\documentclass[letterpaper]{article}
\usepackage[T1]{fontenc}

\usepackage{geometry}
\usepackage{setspace}
\usepackage[
    style=chem-acs,
    articletitle=true, % Forces article titles to be printed
    doi=false,         % Hides DOI (optional, to match your exact example)
    url=false,         % Hides URL (optional)
    eprint=false       % Hides arXiv/eprint (optional)
]{biblatex}

\DeclareFieldFormat{labelnumberwidth}{(#1)}

\usepackage{graphicx}
\usepackage{float}
\newfloat{scheme}{htbp}{los}
\floatname{scheme}{Scheme}
\floatname{chart}{Chart}
\newfloat{graph}{htbp}{loh}

\usepackage{chemformula} % Formulas using \ch{}
\usepackage[version = 4]{mhchem} % Formulas using \ce{}

\usepackage{authblk}
\author{Lorenzo Pettorosso}
\author{Michael Horodynski}
\author{Dmitry K. Polyushkin}
\author{Thomas Mueller}
\affil{%
Technische Universität Wien, Institute of Photonics, Gußhausstraße 27, 1040 Vienna, Austria\\
}

\title{Photonic Crystal Integrated Graphene-Silicon Schottky Diodes for Resonant Infrared Photodetection}
\date{Email: lorenzo.pettorosso@tuwien.ac.at}
\begin{document}

\maketitle

\begin{abstract}
%Narrowband infrared (IR) photodetection has become a critical technology in spectral monitoring and low-noise optical telecommunications. While silicon remains intrinsically transparent to IR radiation, the seamless integration of IR photodetectors with silicon complementary metal-oxide-semiconductor (CMOS) technologies is crucial to realize scalable and cost-effective single-chip architectures. Thanks to its gapless nature and excellent electrical properties, graphene is an ideal candidate for native CMOS integration. In this work, we present a device based on a graphene-silicon Schottky junction with narrowband response in the short-wave IR (SWIR) regime. Integrating a photonic crystal (PhC) slab in the device enables a wavelength-dependent tailoring of graphene absorption. This approach amplifies the device photocurrent in the IR regime by more than two orders of magnitude (> 100×) compared to a reference Gr-Si device without PhC. The resulting enhancement peaks are narrow, and their intensity, number and position can be predicted by numerical simulation of different geometries of PhC slabs. This device’s design offers a highly scalable, CMOS-compatible architecture for narrowband IR photodetection, paving the way for next-generation IR sensors.

%Narrowband infrared (IR) photodetection has become a critical technology in spectral monitoring and low-noise optical telecommunications. It is also highly valuable for In-sensor computing devices in smart photdetectors.
Narrowband infrared (IR) photodetection has become a critical technology for the next generation of smart low-power photodetectors.
%While silicon remains intrinsically transparent to IR radiation, the seamless integration of IR photodetectors with silicon complementary metal-oxide-semiconductor (CMOS) technologies is crucial to realize scalable and cost-effective single-chip architectures.
In this work, we present a device based on a graphene-silicon Schottky junction with narrowband response in the short-wave IR (SWIR) regime. Integrating a photonic crystal (PhC) slab in the device enables a wavelength dependant tailoring of graphene absorption. The devices were characterized by photocurrent measurement, which was enhanced more than two orders of magnitude by the PhC integration. The sharp enhancements obtained can also be predicted by numerical simulation. %These highly selective and tunable photodetectors  low-noise telecommunications. 
This approach offers a highly scalable, tunable, CMOS-compatible architecture for narrowband IR photodetection, paving the way for next-generation, low-power IR sensors.

%Nonetheless, graphene's main drawback is its weak intrinsic per-layer absorption of 2.3\%, which severely limits device performance and limits its application to high-sensitivity photodetection. In this work, we overcome the natural limitation of graphene by presenting a photodetector design based on a graphene-silicon (Gr-Si) Schottky junction with spectrally tunable narrow band photoresponsivity in the short-wave infrared regime. Integrating a Photonic Crystal (PhC) on the silicon side of the junction enables wavelength-dependent tailoring of graphene absorption. This approach amplifies photocurrent in the IR regime by more than two orders of magnitude ($>100\times$) relative to a reference Gr-Si device without PhC, paving the way for highly scalable, cost-effective and CMOS-compatible IR photodetection. 
\end{abstract}

%Integrating infrared (IR) photodetectors with silicon CMOS technologies is crucial for scalable single-chip architectures. However, silicon's 1.1 eV bandgap fundamentally limits IR detection. Graphene is an ideal candidate for native CMOS integration due to its broadband absorption and excellent electrical properties; nonetheless, its weak intrinsic absorption of 2.3\% per layer restricts high-sensitivity applications. To overcome this limitation, this work presents a graphene-silicon (Gr-Si) Schottky junction photodetector with spectrally tunable, narrow photoresponsivity in the short-wave infrared regime. Integrating a Photonic Crystal (PhC) on the silicon side of the junction enables the tailoring of graphene absorption in a wavelength-dependent manner. This approach amplifies IR light absorption by more than two orders of magnitude ($>100\times$) compared to a reference Gr-Si device without the PhC, paving the way for scalable and cost-effective CMOS-compatible IR photodetection.

%\section*{Keywords}
%Infrared, Graphene, Narrowband, Photodetector, Photonic crystal
%\section*{Abbreviations}

\begin{center}
    \includegraphics[]{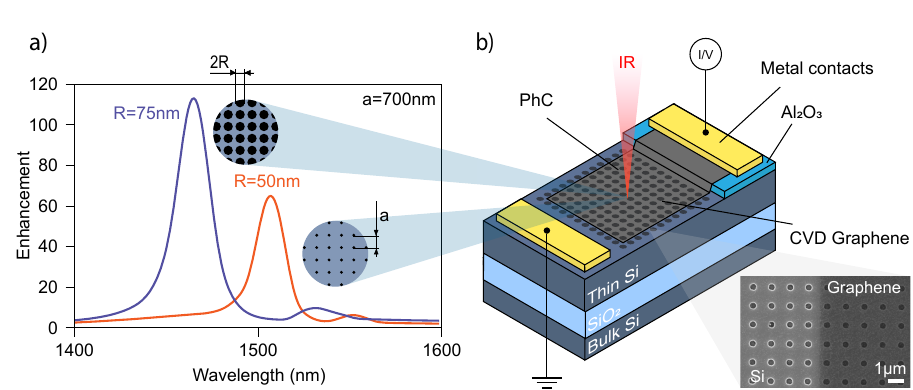}

    \makeatletter
    \def\@captype{figure}
    \makeatother
    
    \caption{a) Spectrum of the photocurrent enhancement compared to a bare Gr-Si device due to a photonic crystal (PhC) with unit cell size $a$=700nm and hole radii ($R$) of 75nm (blue) and 50nm (orange), respectively. b) Schematic of the device design: Graphene sheet on top of the 220nm thin Si PhC slab residing on a 2\textmu m SiO2 layer. The inset shows a scanning electron microscope image of the graphene sheet on the PhC slab.}
    \label{fig:1}
\end{center}
The infrared (IR) spectrum holds relevant information for a wide range of scientific and industrial applications. In recent years, there has been a massive surge in demand for IR sensors across fields such as autonomous vehicles, consumer electronics, biosensing and telecommunications \cite{hashemi_review_2022,li_silicon_2025,butt_midinfrared_2026,zhuge_nanostructured_2017}. The move towards higher data transfer bandwidth makes the development of IR photodetectors (PDs) highly valuable. Specifically, the O-band (1310nm) and C-band (1550nm) for optical telecommunication reside in the short-wave IR (SWIR) regime \cite{wu_shortwave_2023, bruns_next-generation_2017,crisci_unveiling_2025}. Narrowband PDs enable more efficient data transfer, suppressing signal noise and improving channel fidelity. Furthermore, single-wavelength detection is essential in environmental monitoring, as it allows the detection of characteristic molecular absorption bands. 
% as many organic molecules have absorption fingerprints in the near-IR and 
Moreover, narrowband PDs have already proved to be highly valuable for on-chip spectroscopy in smart sensing devices \cite{kwak_-sensor_2023}. Single-wavelength tunable detection is the critical requirement for in-sensor computing devices. This technology can help to reduce computational load and power consumption in the next generation of PDs.
%citations and more info about why narrow detection is nice
Silicon (Si) holds an unrivalled position within the semiconductor industry, making the direct integration of IR-sensitive technologies into silicon-based architectures highly valuable for scalable optoelectronics. However, its bandgap of $1.1\text{ eV}$, which renders it transparent to infrared radiation, remains a fundamental limitation \cite{sze_physics_2006}. Currently, the fabrication of Si-based IR photodetectors relies on heterogeneous integration and monolithic epitaxial growth of narrow bandgap materials, such as III-V semiconductors, on silicon platforms \cite{roelkens_iii-v--silicon_2015, zeng_recent_2023}. Therefore, the development of a CMOS-compatible narrow-band photodetector remains highly valuable. %Consequently, the realization of fully CMOS-integrated IR photodetectors represents a relevant scientific challenge.
Graphene (Gr) is an ideal IR-sensitive candidate for seamless integration into existing CMOS architectures \cite{bonaccorso_graphene_2010,novoselov_electric_2004, jalali_silicon_2006,koppens_photodetectors_2014, romagnoli_graphene-based_2018}. Thanks to its gapless semimetal nature, it offers uniform broadband absorption over a wide spectral range and an ultrafast response owing to its ultrahigh carrier mobility \cite{novoselov_electric_2004,bolotin_ultrahigh_2008}. However, the absorption of pristine single-layer graphene is fundamentally limited to $\pi\alpha \approx 2.3\%$ (where $\alpha$ is the fine structure constant), %(The optical absorption coefficient of single-layer graphene is −ln(1−πα)/d ≈ 7 × 105 cm−1 , independent of wavelength At the technologically important wavelengths of 850, 1300, and 1550 nm, this value is between 1 and 3 orders of magnitude higher than that of conventionally used semiconductor materials such as In0.53Ga0.47As, GaAs, or Ge.21 Nevertheless, due to the short interaction length, a layer of graphene absorbs only πα = 2.3% of the incident light.20 )%
making it practically challenging to use in applications requiring high sensitivity \cite{nair_fine_2008}. To overcome the intrinsic optical limitations of graphene, various strategies have been explored to improve its performance for photodetection applications \cite{pospischil_cmos-compatible_2013,gan_chip-integrated_2013,riazimehr_high_2019,pelella_graphenesilicon_2021}.

%fine position for this \cite{ mueller_graphene_2010}.
%\cite{bonaccorso_graphene_2010}
However, developing PDs with tunable narrow resonances remains challenging.
% 
%While many approaches aim to enhance broadband performance using waveguides \cite{pospischil_cmos-compatible_2013,gan_chip-integrated_2013} or electrical barrier tuning \cite{pelella_graphenesilicon_2021},approaches to generate a narrowband graphene absorption have been explored. 
Although metallic plasmonic nanostructures can provide spectrally localized increases in photoresponsivity, the resonances obtained through these techniques are typically broad and severely hindered by Ohmic losses \cite{wang_enhancing_2020, baranov_all-dielectric_2017, khurgin_how_2015}. The integration of graphene channels into Fabry-Pérot nanocavity resonators has been shown to yield strong, narrow resonances. Nevertheless, this approach relies on the complex fabrication of multilayer distributed Bragg reflectors, where the resonant wavelength is strictly fixed by the cavity's physical dimensions. Consequently, arbitrarily tuning its spectral position remains highly challenging, rendering its implementation in in-sensor computing impractical \cite{furchi_microcavity-integrated_2012}. Furthermore, the development of highly selective Gr-based IR photodetectors remains largely underexplored, as the bandgap of conventional host semiconductors restricts sub-bandgap absorption entirely to the weak intrinsic response of Gr. %Gr-Si photodetectors exhibit high responsivity in the visible regime, where light absorption is governed by silicon \cite{an_tunable_2013}. In the IR regime, the photon energy falls below the Si bandgap.

%When hot carriers are photogenerated in graphene and possess sufficient momentum and energy, they can overcome the barrier and be subsequently injected into the silicon conduction or valence band, generating a photocurrent in the diode. The photodetection performance in the IPE  drastically  as for efficient hot carriers inkectiont trough the schottky junction, the charges must satisfy momentum conditions and energy simultaneusly. the number of avialbale carriers is limited due to the ultrafast hot carrier thermalization in graphene.\cite{brida_ultrafast_2013, goykhman_-chip_2016}.

%they use surface plasmons or nanocavity which are not ideal 

%The low absorption in graphene severely limits the quantum efficiency of such devices in IR regimes.  

Here, we present a CMOS-compatible and highly integrable Gr-Si architecture designed to engineer the device's spectral photoresponsivity to achieve a narrowband response in the short-wave infrared (SWIR) regime while simultaneously overcoming the weak intrinsic graphene absorption.  Our device offers a predictable, controllable, sharp enhancement of photocurrent by integrating a photonic crystal (PhC) on the silicon side of a Gr-Si Schottky diode, using a standard commercial silicon-on-insulator (SOI) wafer. This simplicity is especially valuable for in-sensor computing, which requires a number of resonances at different wavelengths, as these can all be defined in a single lithographic run.
% we present a CMOS-compatible, highly integrable method to engineer the graphene absorption spectrum in the short-wave IR regime by integration of a Photonic crystal into a Grapehene-Silicon (Gr-Si) schottky diode.
PhCs have been extensively studied for their ability to trap light and support guided resonances. The integration of a PhC into a Gr-Si device makes it possible to confine resonant modes on its surface and enhance graphene absorption in a strongly wavelength-dependent manner \cite{piper_broadband_2016,piper_broadband_2016,fan_analysis_2002,rituraj_highly_2025,gu_photonic_2015}.
%, matching the 2D nature of graphene, without introducing ohmic losses due to plasmonic structures and offering a much easier integration compared to plasmonic cavities \cite{baranov_all-dielectric_2017, khurgin_how_2015}.
Figure \ref{fig:1}a illustrates the narrow enhancement from integrating a PhC into a Gr-Si Schottky junction. We measured the output photocurrent at the PhC resonant wavelength to be more than two orders of magnitude ($>100\times$) higher than that of a bare Gr-Si device. As conceptually represented in Figure \ref{fig:1}b, the PhC under study consists of a square lattice arrangement of circular nanoholes in a thin Si slab residing on a SiO2 layer from an SOI wafer. Furthermore, the spectral position of the device's resonant modes can be tuned by modifying the geometrical parameters of the PhC: in Figure \ref{fig:1}a, the two photocurrent enhancement peaks correspond to different PhC hole radii ($R$) with a fixed unit cell size ($a$).

%The device studied in this work is based on a Gr-Si Schottky junction. To overcome the fundamental absorption limit of graphene, a PhC was integrated in the 220nm Si slab which a CVD graphene sheet was wet transferred and shaped as depicted in FIG 1A. When a PhC is illuminated it can confine The PhC studied in this work are composed of a period array of air holes forming a 2D square lattice with a unit cell size from 550nm to 1150 nm. The studied holes' radius was from 25nm to 150nm. The PhC can confine allowed light modes  Such devices were characterised by measuring their spectral responsivity in the 1200nm to 2000nm range. The geometrical parameter of the 2D PhC lattice can be tuned to control the spectral graphene absorption, showing a narrow responsivity peak around the resonant frequency of the PhC. The enhancement effect of the integration of the PhC in a Gr-Si Schottky junction is shown in FIG 1B, for a defined PhC unit cell size and different hole radii.
The PhC-integrated Gr-Si photodetectors were fabricated using standard nanofabrication techniques.
A SOI wafer with a 220 nm p-doped device layer and 2 \textmu m SiO2 insulating layer was covered with 60 nm Al2O3 by atomic layer deposition. 250 \textmu{}m\textsuperscript{2} windows were wet-etched using buffered HF (BOE), exposing the silicon in the device region of the SOI. The photonic crystals were patterned by electron beam lithography (EBL) and etched by an SF$_6$/Ar reactive ion etching (RIE) plasma. Graphene grown by chemical vapour deposition (CVD) was wet transferred and shaped using EBL and O2-plasma RIE. The contacts were deposited by evaporation of 5 nm Ti and 50 nm Au.  %The PhC under study was made by a square lattice displacement of circular nanoholes, defined by geometrical parameters of unit cell size (a) defined as the distance between the holes, and the hole radii (R).
The photocurrent spectrum of the devices was characterized by normal incidence illumination of the Gr-Si junction using a supercontinuum laser with a wavelength resolution of 1nm. The laser was focused on the device to a beam spot diameter of $\approx 30$ \textmu m with an incident power on the order of 300 \textmu W at 1550 nm, the full power spectrum is available in Figure S2. The signal was modulated at 20 Hz and measured using a lock-in amplifier with a 1 s integration time. The bias voltage ($V_b$) was applied by means of a transimpedance amplifier to bring the device into the reverse-biased state. A more detailed description of the measurement setup is available in Figure S3.
%When graphene is brought into contact with a semiconductor like Si, a charge transfer is induced until the Fermi level is aligned between the two sides of the junction. This alignment results in the bending of the valence and conduction bands of the silicon with the relative formation of a Schottky barrier ($\Phi_B$) at the Gr-Si interface. Thanks to the built-in electric field localized at the junction, photogenerated carriers can be separated. Such a device can operate via two different mechanisms depending on the energy of the interacting photons ($h\nu$). For photons whose energy exceeds the silicon bandgap ($h\nu > E_{g,Si}$), electron-hole pairs are generated and separated by the built-in electric field. For photons with energies lower than the bandgap of silicon ($h\nu < E_{g,Si}$), the device operates in the internal photoemission mode (IPM), where the photons are absorbed by the graphene layer, generating hot carriers. When the hot carriers possess sufficient momentum and energy, they can overcome the barrier ($\Phi_B$) and be subsequently injected into the silicon conduction or valence band, generating a photocurrent in the diode. %Since silicon has a bandgap of 1.1 eV, an IR photodetector based on the Gr-Si diode concept inherently operates in the IPM.
\begin{figure}
\centering
\includegraphics{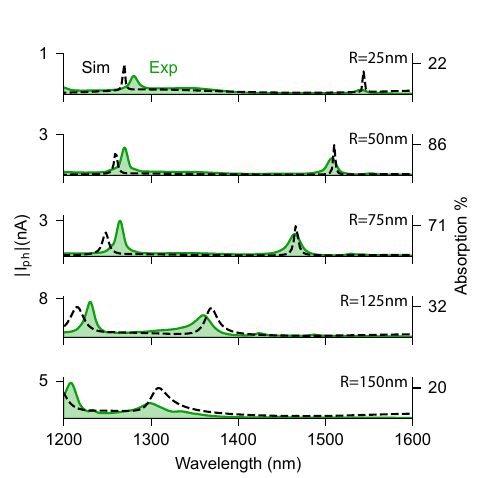}
\caption{Comparison between simulated graphene absorption and measured photocurrent. Photocurrent spectra measured at bias voltage $V_b$ = -1V of a PhC-integrated Gr-Si Schottky diode photodetector with a unit cell size of a=550 nm and different hole radii.}
\label{fig:2}
\end{figure}

The inclusion of a PhC into a Gr-Si photodetector allows us to engineer the device's spectral photoresponsivity in a straightforward manner. To design a device with a specific spectral position of the PhC resonant modes, we simulate the absorption spectrum of the graphene layer with the finite-difference time-domain method, using an open-source software package \cite{oskooi_meep_2010}. We represent the graphene layer in our simulations by a one-pixel-thick conductor with the refractive index of graphene scaled by the simulation's resolution.
%open-source 3D vectorial Maxwell equations solver MEEP \cite{oskooi_meep_2010}. %Then, the device's photoresponsivity was measured.
In Figure \ref{fig:2}, we show measured photocurrent spectra and compare them to the simulated absorption in graphene of a typical device with geometrical parameters $a$=550 nm and $R$ ranging from 25 to 150 nm. As predicted by the simulation, two narrow resonant peaks appear in the 1200–1600 nm range. The photocurrent peak positions match the simulated absorption spectra, with some deviation due to fabrication tolerance discrepancies. The scanning electron microscope images of the PhCs are available in Figure S1. A blueshift of the resonant peaks is observed for both the simulated absorption and the measured photocurrent spectra when increasing the PhC hole radii. This occurs because larger holes increase the air filling factor of the Si slab, reducing its effective refractive index \cite{fan_analysis_2002,joannopoulos_photonic_2011}.
The effective maximum responsivity of this device was measured to be 18 \textmu A/W. The laser power spectrum is available in Figure S2.

In the IR regime, the Gr-Si device operation is attributed to the internal photoemission mechanism, where photon absorption occurs in graphene, and the resulting photoexcited hot carriers can be injected into the Si across the Schottky barrier. For effective injection across the Schottky barrier, hot carriers must simultaneously satisfy the correct energy and momentum conditions. Ultrafast thermalization in graphene further reduces the population of carriers that can inject into silicon, limiting the photocurrent generation\cite{brida_ultrafast_2013, goykhman_-chip_2016}.
Previous works predicted 100\% absorption occurring for critical coupling in Gr-Si devices \cite{piper_total_2014, piper_broadband_2016}. Critical coupling conditions require rate matching in open resonators: the absorption in the graphene layer must exactly balance the radiative decay rate of the cavity \cite{wonjoo_suh_temporal_2004}. Therefore, total absorption cannot be achieved with the current device's design, as the absence of a highly reflective bottom layer leaves the transmission channel open.
The integration of a back reflector into the architecture introduces substantial fabrication complexity.
According to the simulation data, with this combination of geometrical parameters, it is possible to achieve a maximum theoretical absorption of 86\% in the graphene layer, corresponding to a PhC slab with $a$=550 nm and $R$=50 nm.

\begin{figure*}
\centering
\includegraphics[]{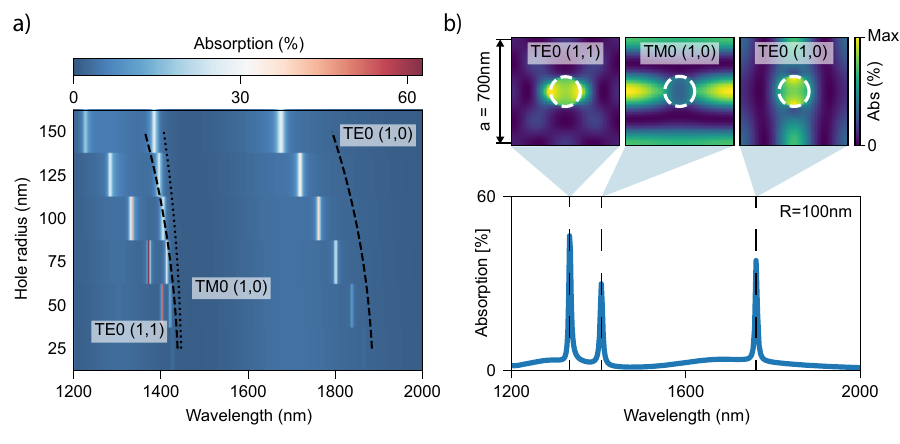}
\caption{a) Simulated absorption spectra in graphene as a function of hole radius ( unit cell size a=700 nm). The dotted line represents an approximate prediction of the resonant modes obtained using Bragg diffraction theory, along with their mode assignments. b) Absorption spectrum of the $R$=100 nm device with the spatial absorption distribution in the Gr layer of the three resonances on top (full unit cell is shown). }
\label{fig:3}
\end{figure*}

The PhC's geometric parameters affect the spatial modes' positions, numbers, widths, and intensities. PhCs with larger unit cell sizes offer a greater number of resonant modes in the same spectral range compared to PhCs with smaller unit cell sizes. The position of the resonant mode for a given PhC can also be approximately predicted by Bragg phase-matching condition grating-coupling:
\begin{equation}
    \lambda_{res} =   \frac{n_{eff}(\lambda)a}{\sqrt{m^2 + n^2}},
    \label{eq:1}
\end{equation}
where $n_{eff}(\lambda)$ is the effective refractive index for light propagating in the silicon slab, $a$ is the PhC unit cell size, and $m$ and $n$ (with $m,n = 0,1,2, (m,n) \neq (0,0)$),  are the diffraction orders. The effective refractive index of the Si slab is calculated using the transverse resonance condition weighted with the hole filling factor \cite{saleh_fundamentals_1992}. The resonant modes predicted by solving Maxwell's equations and their approximate prediction by Bragg's diffraction law are schematically compared in Figure \ref{fig:3}a. The Bragg's diffraction approximation works well for smaller filling factors, after which the structure deviates more from the effective medium theory \cite{lalanne_effective_1996, choy_effective_2015}.

%\begin{equation}
 %   2k_{0}n_{1}h \cos \theta_{1}^{(m)} - \phi_{u} - \phi_{l} = 2m\pi
%\end{equation}
%\begin{equation}
 %   \phi_{TE} = 2 \tan^{-1}\frac{\sqrt{\sin^{2}\theta^{(m)}_{1} - (n_{2}/n_{1})^{2}}}{\cos \theta^{(m)}_{1}}
%\end{equation}
%\begin{equation}
 %   \phi_{TM} = 2 \tan^{-1}\frac{\sqrt{(n_{1}/n_{2})^{2}\sin^{2}\theta^{(m)}_{1} - 1}}{(n_{2}/n_{1})\cos \theta^{(m)}_{1}}
%\end{equation}
%\begin{equation}
 %  n^{(m)}_{eff}= n_1\sin\theta^{(m)}_{1}
%\end{equation}
%trancendental equation solved numerically. Describe where the wave are standing so people undrstand we are solving neff for the Si slab
%However, it was found that the analogy between subwavelength periodic structures and homogeneous media is less accurate for 2-D than for 1-D structures. For periods larger than half the cutoff period, the periodic structure of our example stops behaving as a homogeneous medium
%explain code what is doing in practice to get the resonant modes.
%why better matching for smaller holes?
Our simulations also reveal that each absorption resonance is associated with a specific spatial absorption distribution in the Gr layer. The silicon PhC sits between air and a 2 \textmu m SiO2 dielectric layer (Figure \ref{fig:1}b), making the TE and TM resonant modes non-degenerate due to the refractive index mismatch. Each resonant peak can be attributed to the excitation of fundamental and higher-order guided resonances within the PhC slab. Furthermore, the Bragg condition defines the phase-matching criterion under which the periodic lattice strongly scatters the guided light, generating localized resonances. In Figure \ref{fig:3}b, the resonant modes calculated using the Bragg condition are correlated with the absorption peaks, and the corresponding spatial absorption distributions are simulated by solving Maxwell's equations. A clear correlation between the shape of the spatial distribution and the diffraction order of the Bragg resonant modes is evident. From this, it is possible to conclude that each absorption peak can be attributed to the excitation of fundamental and higher-order guided resonances within the PhC slab, depending on its geometry. In the example reported in Figure \ref{fig:3}b, the first peak is associated with a TE0 (1,1) mode that propagates along the diagonal direction of the square lattice. The second peak is defined by a TM0 (1,0) mode propagating along the y-axis, and the third peak corresponds to a TE0 (1,0) mode propagating along the x-axis.
%IF NO MILLER INDICES: In the example reported in Figure \ref{fig:3}b, the first peak is associated with a TE0 (1,1) mode that propagates along the diagonal of the unit cell. The second peak is defined by a TE0 (1,0) mode propagating along the y-axis, and the third peak corresponds to a TM0 (1,0) mode propagating along the orthogonal x-axis.
\begin{figure*}
\centering
\includegraphics{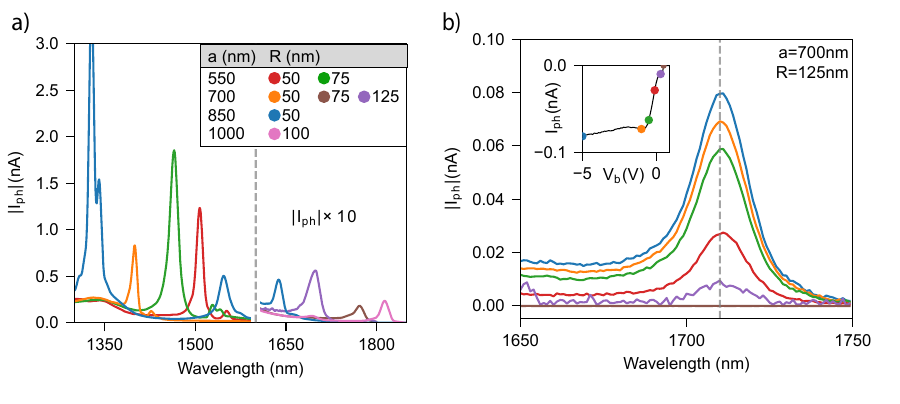}
\caption{a) Photocurrent spectra of PhC-integrated Gr-Si Schottky diode devices with different geometrical parameters, measured at $V_b = -1 \text{V}$. b) Peak photocurrent at 1710 nm for a Gr-Si Schottky diode with PhC geometrical parameters of $a$=700 nm and $R$=125 nm for different $V_b$.}
\label{fig:4}
\end{figure*}

The entire spectral range of interest can be covered by resonant photocurrent-enhancement peaks by selecting appropriate geometrical parameters ($R, a$) for the integrated PhC in the Gr-Si diodes. We show in Figure \ref{fig:4}a a collection of PhC-integrated Gr-Si photodetectors and their spectra that cover a broad wavelength range, with isolated resonances characterized by a full width at half maximum (FWHM) around 20nm. In addition to the resonance tunability, the device's photocurrent output can be dynamically controlled by means of the bias voltage $V_b$ (Figure \ref{fig:4}b). Under reverse bias, the photocurrent increases linearly with $V_b$ down to $V_b$ = -1 V, after which the extraction efficiency approaches the maximum, and the photocurrent remains limited by the incident optical generation rate. The residual modest increase in the photocurrent can be attributed to the voltage-induced lowering of the Schottky barrier\cite{di_bartolomeo_graphene_2016, sze_physics_2006}. The ability to tune the photoresponse is a crucial requirement for developing low-power %smart
photodetectors, as they operate effectively within a narrow voltage window, thereby limiting power consumption.

In conclusion, we demonstrated that the photoresponse of a Gr-Si Schottky diode photodetector can be effectively tuned by integrating a PhC. By optimizing the geometric parameters of the PhC, specifically the unit cell size and air-hole radii, resonant peaks were successfully engineered at targeted spectral positions within the SWIR regime. Furthermore, a close correspondence between the simulated absorption resonances and the measured photocurrent peak positions was observed. This validates the use of numerical simulations as a robust predictive tool for engineering device performance. Crucially, the PhC integration yielded a photocurrent enhancement of more than two orders of magnitude ($>100\times$) compared to a planar Gr-Si reference device. Specifically, the PhC integration results in sharp, narrowband resonances in the SWIR regime with FWHM below 20nm, making the device useful for low-noise, high-precision telecommunications. By careful tuning of the geometrical parameters in a series of Gr-Si devices, we were able to cover the spectral range under study with narrow photocurrent enhancement peaks, making this design highly valuable for in-sensor computing smart photodetectors and spectroscopy applications.

While tuning the 2D lattice parameters proved sufficient to control the spectral response, future optimizations could explore additional degrees of freedom, such as bias voltage, silicon slab thickness, and alternative PhC cavity designs. Controlling more parameters can further maximize the optical field enhancement within the graphene layer. Although the output photocurrent generated by the fabricated devices aligns with that of similar Gr-Si IR photodetectors \cite{casalino_vertically_2017}, the absolute photoresponsivity remains limited. This may partially originate from the formation of an interfacial native SiO2 layer, hindering the injection of hot electrons across the Schottky junction. Nevertheless, the primary focus of this study was to isolate and validate the optical enhancement achieved through PhC integration. Addressing these interface limitations in future fabrication processes, combined with the PhC design principles presented in this work, promises to unlock high performance for advanced IR, compatible with silicon-based device architectures.
%In conclusion, we have demonstrated it is possible to tune the responsivity of a Gr-Si schottky diode photodetector by mean of a PhC integration. By selecting the geometrical parameters of the PhC it was possible to generate a resonance peak to a specific spectral position in the SWIR regime. We demonstrated that the graphene absoprtion, simulated by solving Maxwell's equation, is related to the photoresponsivity of the device. Thus, numerical simulations can be use to predict and engineer the device performance. The enhancment effect of the PhC in the Gr-Si photodetector was measured to be larger than two order of magnitude compared to a plain device. In this study, the PhC was charachterised by mainly the air hole radii and the unit cell size of the lattice. While tuning these parameters was enoughf to control the spectral responsivity in the device, other parameters such as the bias voltage, the Si slab thickness, the PhC nanostrucre shape can be change to modify the absoprtion spectrum ehnancment induced in the grapheen layer. While the output photocurrent measured in our devices is in the same order of magnitude compared to similar IR Gr-Si based photodetectors, the measured photoresponsivity is not satisfactory. This could be related to a low quality Gr-Si interface: the formation of a native SiO2 layer would limit hot electrons to cross the junction decreasing performance. The suty focus hereby on the PhC enhancment effect compared to a plain Gr-Si Schottky diode photodetector in the IR regime.

\section*{Acknowledgements}

This research was funded in part by the Austrian
Science Fund (FWF) [10.55776/DOC142] (TU-DX: Towards Applications of 2D Materials). This research was funded in part (M.H.) by the Austrian Science Fund (FWF) [10.55776/J4729].

%\section*{Supporting information}

%A listing of the contents of each file supplied as Supporting Informationshould be included. For instructions on what should be included in the Supporting Information as well as how to prepare this material for publications, refer to the journal's Instructions for Authors.

\printbibliography

\end{document}